\documentclass[11pt]{article}
\usepackage[letterpaper]{geometry}

\newcommand{\idlow}[1]{\mathord{\mathcode`\-="702D\it #1\mathcode`\-="2200}}
\newcommand{\id}[1]{\ensuremath{\idlow{#1}}}
\newcommand{\litlow}[1]{\mathord{\mathcode`\-="702D\sf #1\mathcode`\-="2200}}
\newcommand{\lit}[1]{\ensuremath{\litlow{#1}}}

\usepackage{graphicx}
\usepackage{multirow}
\usepackage{wrapfig}
\usepackage{hyperref}

\date{}

\title{Towards Reduced Instruction Sets for Synchronization}

\author{
  Rati Gelashvili\\
  \small MIT\\
  \texttt{gelash@mit.edu}
\and 
  Idit Keidar\\
  \small Technion\\
  \texttt{idish@ee.technion.ac.il}
\and
  Alexander Spiegelman\\
  \small Technion\\
  \texttt{sashas@tx.technion.ac.il}
\and
  Roger Wattenhofer\\
  \small ETH Zurich\\
  \texttt{wattenhofer@ethz.ch}
}

\begin{document}

\maketitle
\begin{abstract}
% {
% \small
Contrary to common belief, a recent work by Ellen, Gelashvili, Shavit, and Zhu 
  has shown that computability does not require multicore architectures to support 
  ``strong''  synchronization instructions like \id{compare-and-swap}, as opposed 
  to combinations of ``weaker'' instructions like \id{decrement} and \id{multiply}.
However, this is the status quo, and in turn, most efficient 
  concurrent data-structures heavily rely on \id{compare-and-swap} 
  (e.g. for swinging pointers and in general, conflict resolution).

We show that this need not be the case, by designing and implementing 
  a concurrent linearizable \lit{Log} data-structure
  (also known as a \lit{History} object), 
  supporting two operations: \id{append(item)}, which appends the item to the log, 
  and \id{get-log()}, which returns the appended items so far, in order. 
Readers are wait-free and writers are lock-free, and this
  data-structure can be used in a lock-free universal construction to
  implement any concurrent object with a given sequential specification.
Our implementation uses atomic \id{read}, \id{xor}, \id{decrement}, and 
  \id{fetch-and-increment} instructions supported on X86 architectures, 
  and provides similar performance to a \id{compare-and-swap}-based solution
  on today's hardware.
This raises a fundamental question about minimal set of synchronization instructions 
  that the architectures have to support.
% }
\end{abstract}

\section{Introduction}
In order to develop efficient concurrent algorithms and data-structures in multiprocessor systems,
  processes that take steps asynchronously need to coordinate their actions.
In shared memory systems, this is accomplished by applying hardware-supported low-level 
  atomic instructions to memory locations.
An atomic instruction takes effect as a single indivisible step.
The most natural and universally supported instructions are \id{read} and \id{write},
  as these are useful even in uniprocessors to store and load data from memory.

A concurrent implementation is \emph{wait-free}, if any process that takes infinitely 
  many steps completes infinitely many operation invocations.
An implementation is \emph{lock-free} if in any infinite execution infinitely
  many operations are completed.
The celebrated FLP impossibility result~\cite{FLP85}
  implies that in a system equipped with only \id{read} and \id{write} instructions, 
  there is no deterministic algorithm to solve binary lock-free/wait-free consensus 
  among $n \geq 2$ processes.
Binary consensus is a synchronization task where processes start with input bits,
  and must agree on an output bit that was an input to one of the processes.
For one-shot tasks like consensus, wait-freedom and lock-freedom are equivalent.

Herlihy's Consensus Hierarchy~\cite{Her91} takes the FLP result further.
% Herlihy's Consensus Hierarchy~\cite{Her91} assigns a \emph{consensus number} to each object,
It assigns a \emph{consensus number} to each object, namely, the number of processes for which
  % namely, the number of processes for which 
  there is a wait-free binary consensus algorithm using only instances of this object 
  and \id{read}-\id{write} registers.
An object with a higher consensus number is hence a more powerful tool for synchronization.
Moreover, Herlihy showed that consensus is a fundamental synchronization task, 
  by developing a universal construction which allows $n$ processes to wait-free implement 
  any object with a sequential specification, provided that they can solve consensus among themselves.

Herlihy's hierarchy is simple, elegant and, for many years, has been our best explanation 
  of synchronization power.
It provides an intuitive explanation as to why, for instance, the \id{compare-and-swap} 
% Herlihy's hierarchy provides an explanation as to why, for instance,
  instuction can be viewed ``stronger'' than \id{fetch-and-increment}, 
  as the consensus number of a \lit{Compare-and-Swap} object is $n$,
  while the consensus number of \lit{Fetch-and-Increment} is $2$.
  
However, key to this hierarchy is treating synchronization instructions as distinct objects, 
  an approach that is far from the real-world, where multiprocessors do let processes apply
  supported atomic instructions to arbitrary memory locations. 
In fact, a recent work by Ellen et al.~\cite{EGSZ16} has shown that a combination 
  of instructions like \id{decrement} and \id{multiply-by-n}, whose corresponding objects 
  have consensus number $1$ in Herlihy's hierarchy, when applied to the same memory location, 
  allows solving wait-free consensus for $n$ processes.
Thus, in terms of computability, a combination of instructions traditionally viewed as ``weak''
  can be as powerful as a \id{compare-and-swap} instruction, for instance.

The practical question is whether we can really replace a \id{compare-and-swap} instruction 
  in concurrent algorithms and data-structures with a combination of weaker instructions.
This might seem improbable for two reasons.
First, 
\id{compare-and-swap} is ubiquitous in practice and used heavily 
  for various tasks like swinging a pointer.
Second,
% Also,  
the protocol given by Ellen et al. solves only binary $n$-process consensus.
It is not clear how to use it for implementing complex concurrent objects,
  as utilizing Herlihy's universal construction is not a practical solution.
On the optimistic side, there exists a concurrent queue implementation based on
  \id{fetch-and-add} that outperforms \id{compare-and-swap}-based alternatives~\cite{MA13}.
Both a \lit{Queue} and a \lit{Fetch-and-Add} object have consensus number $2$,
  and this construction does not ``circumvent'' Herlihy's hierarchy 
  by applying different non-trivial synchronization instructions to the same location.
Indeed, we are not aware of any practical construction that relies on applying different 
  instructions to the same location.

As a proof of concept, we develop a lock-free universal construction using only 
  \id{read}, \id{xor}, \id{decrement}, and \id{fetch-and-increment} instructions.
The construction could be made wait-free by standard helping techniques.
In particular, we implement a \lit{Log} object~\cite{tango} 
  (also known as a \lit{History} object~\cite{matei}),
  which supports high-level operations \id{get-log()} and \id{append(item)},
  and is linearizable~\cite{HW90} to the sequential specification that 
  \id{get-log()} returns all previously appended items in order.
This interface can be used to agree on a simulated object state,
  and thus, provides the universal construction~\cite{Her91}. 
In practice, we require a \id{get-log()} for each thread to return
  a suffix of items after the last \id{get-log}() by this thread.
We design a lock-free \lit{Log} with wait-free readers,
  which performs as well as a \id{compare-and-swap}-based solution on modern hardware.

In our construction, we could replace both \id{fetch-and-increment}
  and \id{decrement} with the atomic \id{fetch-and-add} instruction, 
  reducing the instruction set size even further. 

\section{Algorithm}
\begin{wrapfigure}{l}{0.34\linewidth} 
%    \centering 
   \includegraphics[width=2in]{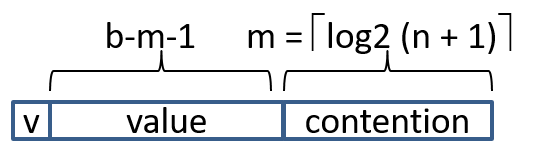}
   \caption{\small Element of $A$.}
    \label{fig:element}
 \end{wrapfigure}

We work in the bounded concurrency model where at most $n$ processes will ever 
  access the \lit{Log} implementation.
The object is implemented by a single \id{fetch-and-increment}-based counter $C$, 
  and an array $A$ of $b$-bit integers on which 
  the hardware supports atomic \id{xor} and \id{decrement} instructions.
We assume that $A$ is unbounded.
Otherwise, processes can use $A$ to agree on the next array $A'$ 
  to continue the construction.
$C$ and the elements of $A$ are initialized by $0$.
We call an array location \emph{invalid} if it contains a negative value,
  i.e., if its most significant bit is $1$,
  \emph{empty} if it contains value $0$, and \emph{valid}
  otherwise.
The least significant $m = \lceil \log_2{(n+1)} \rceil$
bits are \emph{contention bits} and have a special importance to the algorithm.
The remaining $b - m - 1$ bits are used to store items.
See Figure~\ref{fig:element} for illustration.

For every array location, at most one process will ever attempt to record 
  a $(b-m-1)$-bit item, and at most $n-1$ processes will attempt to invalidate this location.
No process will try to record to or invalidate the same location twice.
In order to record item $x$, a process invokes \id{xor}$(x')$, where $x'$ is $x$ shifted by $m$ bits
  to the left, plus $2^m -1 \geq n$, i.e., the contention bits set to $1$.
To invalidate a location, a process calls a \id{decrement}.
The following properties hold:
\begin{itemize}
\item[1.] After a \id{xor} or \id{decrement} is performed on a location, no \id{read} on it ever returns $0$.
\item[2.] If a \id{xor} is performed first, no later read returns an invalid value.
  Ignoring the most significant bit, the next most significant $b-m-1$ bits contain
  the item recorded by \id{xor}.
\item[3.] If a \id{decrement} is performed first, then all values returned by later \id{read}s are invalid.
\end{itemize}
A \id{xor} instruction \emph{fails to record an item} if it is performed after a decrement.

To implement a \id{get-log}$()$ operation, process $p$ starts at index $i = 0$, 
  and keeps reading the values of $A[i]$ and incrementing $i$ until it encounters an empty location $A[i] = 0$.
By the above properties, from every valid location $A[j]$, it can extract the item $x_j$ recorded 
  by a \id{xor}, and it returns an ordered list of all such items $(x_{i_1}, x_{i_2}, \ldots, x_{i_k})$.
In practice, we require $p$ to return only a suffix of items appended after the 
  last \id{get-log}() invocation by $p$.
This can be accomplished by keeping $i$ in static memory instead of initializing it to $0$
  in every invocation.
To make \id{get-log} wait-free, $p$ first performs $l = C.\id{read()}$.
Then, if $i$ becomes equal to $l$ during the traversal, it stops and returns the items extracted so far.

To implement \id{append}$(x)$, process $p$ starts by $\ell = C.\id{fetch-and-increment}()$.
Then it attempts to record item $x$ in $A[\ell]$ using an atomic \id{xor} instruction.
If it fails to record an item, the process does another \id{fetch-and-increment} and 
  attempts \id{xor} at that location, and so on, until it is able to successfully record $x$.
Suppose this location is $A[\ell']$.
Then $p$ iterates from $j=\ell'-1$ down to $j=0$, reading each $A[j]$, and if $A[j]$ is empty,
  performing a \id{decrement} on it.
Afterwards, process $p$ can safely return.

% The proofs of lock-freedom and linearizability can be found in the full version at
  % \url{https://alexanderspiegelman.github.io/XorFaaUniversal.pdf}.
% \footnote{\url{http://webee.technion.ac.il/~idish/ftp/XorFaaUniversal.pdf}}.

\id{fetch-and-increment} guarantees that each location is \id{xor}ed at most once,
  and it can be \id{decrement}ed at most $n-1$ times, once by each process that did not \id{xor}. 
As a practical optimization, each process can store the maximum $\ell'$ from its previous \id{append} operations
  and only iterate down to $\ell'$ in the next invocation (all locations with lower indices will be non-empty).
Our implementation of \id{append} is lock-free, because if an operation takes steps and does not terminate 
  it must be repeatedly failing to record items in locations.
This only happens if other \id{xor} operations successfully record their items and invalidate these locations.

At any time $t$ during the execution, let us denote by $f(t)$ as the maximum index such that,
  $A[f(t)]$ is valid and $A[j]$ is non-empty for all $j \leq f(t)$.
By the first property $f(t)$ is non-decreasing, i.e., for $t' > t$ we have $f(t') \geq f(t)$.
We linearize an \id{append}$(x)$ operation by $p$ that records $x$ at location $A[\ell]$ 
  at the smallest $t$ where $f(t) \geq \ell$.
This happens during the operation by $p$,
  as when $p$ starts \id{append}$(x)$, $A[\ell]$ is empty,
  and when $p$ finishes, $A[0] \neq 0, \ldots, A[\ell-1] \neq 0$ and $A[\ell]$ is valid. 
Next, we show how to linearize \id{get-log}$()$.

Consider a \id{get-log}$()$ operation with the latest returned item $x_{\ell}$ extracted from $A[\ell]$.
We show by contradiction that the execution interval of this \id{get-log}$()$ must contain time $t$ 
  such that $f(t) = \ell$.
We then linearize \id{get-log}$()$ at the smallest such $t$.
It is an easy exercise to deal with the case when multiple operations are linearized at exactly 
  the same point by slightly perturbing linerization points to enforce the correct ordering.
Suppose the \id{get-log}$()$ operation extracts $x_{\ell}$ from $A[\ell]$ at time $T$.
$f(T) \geq \ell$ as \id{get-log}$()$ stops at an empty index, 
  and by the contradiction assumption we must have $\ell' = f(T) > \ell$. 
\id{get-log}$()$ then reaches valid location $A[\ell']$ and extracts an item $x_{\ell'}$ from it, 
  contradicting the definition of $x_{\ell}$.

We implemented the algorithm on X86 processor and with $32$ threads. 
It gave the same performance as
  an implementation that used \id{compare-and-swap} for recording items and invalidating locations.
It turns out that in today's architectures, the cost of supporting \id{compare-and-swap} 
  is not significantly higher than that of supporting \id{xor} or \id{decrement}.
This may or may not be the case in future Processing-in-Memory architectures~\cite{PACFKKTY97}.
Finding a compact set of synchronization instructions that, when supported, is equally powerful 
  as the set of instructions used today is an important question to establish in future research.
  
\bibliographystyle{alpha}
% \footnotesize
\bibliography{biblio}

\newcommand{\etalchar}[1]{$^{#1}$}
\begin{thebibliography}{BMW{\etalchar{+}}13}

\bibitem[BMW{\etalchar{+}}13]{tango}
Mahesh Balakrishnan, Dahlia Malkhi, Ted Wobber, Ming Wu, Vijayan Prabhakaran,
  Michael Wei, John~D Davis, Sriram Rao, Tao Zou, and Aviad Zuck.
\newblock Tango: Distributed data structures over a shared log.
\newblock In {\em Proceedings of the 24th ACM Symposium on Operating Systems
  Principles}, SOSP '13, pages 325--340, 2013.

\bibitem[Dav04]{matei}
Matei David.
\newblock Wait-free linearizable queue implementations, 2004.

\bibitem[EGSZ16]{EGSZ16}
Faith Ellen, Rati Gelashvili, Nir Shavit, and Leqi Zhu.
\newblock A complexity-based hierarchy for multiprocessor
  synchronization:[extended abstract].
\newblock In {\em Proceedings of the 35th ACM Symposium on Principles of
  Distributed Computing}, 2016.

\bibitem[FLP85]{FLP85}
Michael~J Fischer, Nancy~A Lynch, and Michael~S Paterson.
\newblock Impossibility of distributed consensus with one faulty process.
\newblock {\em Journal of the ACM (JACM)}, 32(2):374--382, 1985.

\bibitem[Her91]{Her91}
Maurice Herlihy.
\newblock Wait-free synchronization.
\newblock {\em ACM Transactions on Programming Languages and Systems}, 1991.

\bibitem[HW90]{HW90}
Maurice Herlihy and Jeannette Wing.
\newblock Linearizability: A correctness condition for concurrent objects.
\newblock {\em ACM Transactions on Programming Languages and Systems (TOPLAS)},
  12(3):463--492, 1990.

\bibitem[MA13]{MA13}
Adam Morrison and Yehuda Afek.
\newblock Fast concurrent queues for x86 processors.
\newblock In {\em Proceedings of the 18th ACM SIGPLAN Symposium on Principles
  and Practice of Parallel Programming}, volume~48 of {\em PPoPP '13}, pages
  103--112, 2013.

\bibitem[PAC{\etalchar{+}}97]{PACFKKTY97}
David Patterson, Thomas Anderson, Neal Cardwell, Richard Fromm, Kimberly
  Keeton, Christoforos Kozyrakis, Randi Thomas, and Katherine Yelick.
\newblock A case for intelligent ram.
\newblock {\em IEEE Micro}, 17(2):34--44, 1997.

\end{thebibliography}

\end{document}